\documentclass[review]{elsarticle}
\usepackage{xcolor}
\usepackage{multirow}
\usepackage{tabularx}
\usepackage{amsmath}
\usepackage{lineno}
\usepackage{makecell}
\usepackage{amssymb}
\usepackage{float}

\usepackage{lineno,hyperref}

\journal{Journal of High Energy Astrophysics}

\begin{document}

\begin{frontmatter}

\title{Search for Magnetic Monopoles with ten years of the ANTARES neutrino telescope}

\author[a,b]{A.~Albert}
\author[c]{S.~Alves}
\author[d]{M.~Andr\'e}
\author[e]{M.~Anghinolfi}
\author[f]{G.~Anton}
\author[g]{M.~Ardid}
\author[g]{S.~Ardid}
\author[H]{J.-J.~Aubert}
\author[i]{J.~Aublin}
\author[i]{B.~Baret}
% ...

\address[a]{Universit\'e de Strasbourg, CNRS,  a UMR 7178, F-67000 Strasbourg, France}
\address[b]{\scriptsize Universit\'e de Haute Alsace, F-68100 Mulhouse, France}
\address[c]{IFIC - Instituto de F\'isica Corpuscular (CSIC - Universitat de Val\`encia) c/ Catedr\'atico Jos\'e Beltr\'an, 2 E-46980 Paterna, Valencia, Spain}
\address[d]{Technical University of Catalonia, Laboratory of Applied Bioacoustics, Rambla Exposici\'o, 08800 Vilanova i la Geltr\'u, Barcelona, Spain}
\address[e]{INFN - Sezione di Genova, Via Dodecaneso 33, 16146 Genova, Italy}
\address[f]{Friedrich-Alexander-Universit\"at Erlangen-N\"urnberg, Erlangen Centre for Astroparticle Physics, Erwin-Rommel-Str. 1, 91058 Erlangen, Germany}
\address[g]{Institut d'Investigaci\'o per a la Gesti\'o Integrada de les Zones Costaneres (IGIC) - Universitat Polit\`ecnica de Val\`encia. C/  Paranimf 1, 46730 Gandia, Spain}
\address[H]{Aix Marseille Univ, CNRS/IN2P3, CPPM, Marseille, France}
\address[i]{Universit\'e de Paris, CNRS, Astroparticule et Cosmologie, F-75013 Paris, France}
% ...

\begin{abstract}
This work presents a new search for magnetic monopoles using data taken with the ANTARES neutrino telescope over a period of 10 years (January 2008 to December 2017). Compared to previous ANTARES searches, this analysis uses a run-by-run simulation strategy, with a larger exposure as well as a new simulation of magnetic monopoles taking into account the Kasama, Yang and Goldhaber model for their interaction cross-section with matter. No signal compatible with the passage of relativistic magnetic monopoles is observed, and upper limits on the flux of magnetic monopoles with \textit{$\beta$ = v⁄c} $\geq$ 0.55, are presented. 
For ultra-relativistic magnetic monopoles the flux limit is $\sim$ 7$\times$$10^{-18}$ $\rm cm^{-2} \; s^{-1} \; sr^{-1}$.
\end{abstract}

\begin{keyword}
\texttt ANTARES telescope\sep Magnetic Monopoles\sep Neutrino
\end{keyword}

\end{frontmatter}

\section{Introduction}
\label{sec:intro}
The existence of magnetic charges has been considered since a long time ago. The introduction of hypothetical magnetic charges and magnetic currents can restore the symmetry in Maxwell's equations with respect to magnetic and electric fields. When investigating the symmetry between electricity and magnetism, Paul Dirac proved in 1931 \cite{Dirac:1931kp}, that the introduction of Magnetic Monopoles (MMs) can also elegantly solve the problem of the quantization of electric charge. In addition to this, Grand Unified Theories (GUTs) \cite{Lazarides:1986rt} also predicted that MMs could be created shortly after the Big Bang.

Magnetic monopoles are topologically stable particles and carry a magnetic charge defined as a multiple integer of the Dirac charge :
\begin{equation}
    g_{D}=\frac{\hbar c}{2 e}=\frac{e}{2 \alpha}=68.5 e
\end{equation}

\noindent where $e$ is the electon electric charge, $c$ is the velocity of light in vacuum, $\hbar$ is the Planck constant and $\alpha \simeq 1 / 137$ is the fine structure constant.
While Dirac demonstrated the consistency of MMs with quantum mechanics, G. 't Hooft \cite{Gross:1983hb} and Polyakov \cite{Polyakov:1974ek} proved the necessity of MMs in GUTs. This led to the conclusion that any unification model in which the U(1) subgroup of electromagnetism is embedded in a semi-simple gauge group, which is spontaneously broken by the Higgs mechanism, possesses monopole-like solutions. The masses of MMs can range from $10^8$ to $10^{17}$ ${\rm GeV/c^2}$. Larger MM masses are expected if gravity is brought into the GUT picture, as well as in some supersymmetric models \cite{Patrizii:2015uea}. 
Moreover, MMs would be created after the Big Bang (during the phase transition of symmetry breaking), and they would be accelerated by galactic magnetic fields. The rarity of GUT MMs is also a motivation to the scenario of inflation \cite{Yokoyama:1989xj}.

In this paper, a search based on an ANTARES data set of 2480 days collected from 2008 to 2017 is presented. This analysis improves our previous results \cite{ANTARES:2017qjw}, it is based on a new selection yielding a better separation of the putative MM signal, in different velocity $v=\beta c$ ranges, from the background induced by atmospheric muons and neutrinos. The optimization of the selection uses the Model Rejection Factor method \cite{HILL2003393} which allows for the calculation of the sensitivity (and consequently of the upper limit on the flux) using the Feldman-Cousins \cite{Feldman:1997qc} statistical method. Differently from our previous publication in which the Mott model \cite{Ahlen:1976jw} of cross-section with matter was adopted, in this work, the simulation of MM interaction with matter relies on the Kasama, Yang and Goldhaber (KYG) model  \cite{Kazama:1976fm}.

The paper is organized as follows: a brief description of the ANTARES telescope and the expected signal from magnetic monopoles is given in section 2. The simulation of signal and background is described in section 3, while the trigger logic and reconstruction method is summarized in section 4. The MM-sensitive observables, the selection strategy and the optimization method are presented in section 5. The result of the search, the upper limit calculation and the comparison with other experiments are discussed in section 6.

\section{The ANTARES neutrino telescope and the expected signal from MMs}

 The ANTARES (Astronomy with a Neutrino Telescope and Abyss environmental RESearch) detector \cite{AGERON201111} is a Cherenkov  neutrino telescope, anchored at 2475 m below the surface of the Mediterranean Sea and 40 km offshore from Toulon (France). The detector contains 12 detection lines of about 350 m each, horizontally spaced 60 m to 75 m apart and covering a surface area of about 0.1 km$^2$. Each line has 25 floors with 3 optical modules containing each a 10 inch photomultiplier tube (PMT).
These PMTs (Hamamatsu R7081-20) are sensitive to photons in the wavelength range \textit{$\lambda$} $\sim [300, 600]$ nm. PMT signals with an amplitude higher than the threshold of 0.3 photo-electrons are captured in a time window of 40 ns, digitized and registered as hits \cite{ANTARES:2001nhp}, \cite{ANTARES:2006gto}. After digitization the data is then sent under the form of packages of 104 ms, to an on-shores farm of computers for further data processing and filtering. 
Finally, the data goes through a system of triggers, in order to select signals that may correspond to the passage of relativistic particles.


 The signal of MMs in a neutrino telescope is similar to that of high 
energy muons. Tompkins \cite{PhysRev.138.B248} showed that, similarly to electric charge, magnetically charged particles produce Cherenkov emission when their
velocity is higher than the Cherenkov threshold \textit{$\beta_{th} = 1/n,$} where \textit{$n$} is the phase refractive index of the medium. 
Below Cherenkov threshold, the interaction of MMs with electrons in water produces knock-on electrons (also called delta-rays) that, in turn, induce Cherenkov light. The total number of Cherenkov photons having a wavelength between 300 and 600 nm ($N_{\gamma}$), per unit path length of the monopole ($dx$) is calculated using the Berger formula \cite{SELTZER1984665}, and can be determined by:
\begin{equation}
   \frac{d N_{\gamma}}{d x}=\int_{T_{0}}^{T_{m}} \frac{d^{2} N_{e}}{d T_{e} d x}\left[\int_{T_{0}}^{T_{e}} \frac{d N_{\gamma}}{d x_{e}}\left(\frac{d E_{e}}{d x_{e}}\right)^{-1} d E_{e}\right] d T_{e},
\end{equation}
\noindent The electron can induce light if its kinetic energy {\textit{$T_e$}} is above the threshold \textit{T$_{0}$}=0.25 MeV.
{\textit{$T_{m}$}} is the classical upper limit on the energy that 
can be transferred to an atomic electron in a single collision with a 
MM, $E_e$ is the total energy of the delta-ray, $\frac{d^{2} N_{e}}{d T_{e} d x}$ is the distribution of delta-rays produced by a MM and $dx_e$ is the unit length travelled by an electron.
The maximum energy transfer can be approximated by: 
\begin{equation}
 T_{max} = 2\textit{m}_{e}c^{2}\beta^{2}\gamma
^{2},
\end{equation}
\noindent where $\gamma$ is the Lorentz boost factor. This maximum energy is defined as:
\begin{equation}
    \begin{cases}
{\textit{T$_{m}$}} = 0.69 \times {\textit{T$_{max}$}}  $ \- for the Mott cross-section model, corrected by Ahlen \cite{Ahlen:1976jw}$,\\
{\textit{T$_{m}$=}}{\textit{ T$_{max}$}} $ \- for the KYG  model \cite{Kazama:1976fm}.$
 \end{cases}\,.
\end{equation}
\begin{figure}[H]
\centering
\includegraphics[width=0.9\linewidth]{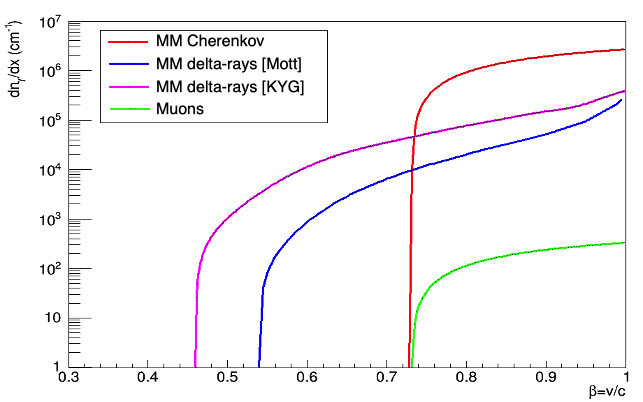}
\caption{Number of Cherenkov photons emitted per cm in the sea water from a  magnetic monopole (red line), and from delta-rays produced along its path according to the Mott model \cite{Ahlen:1976jw} (blue line), and to the KYG model \cite{Kazama:1976fm} (magenta line) as a function of the velocity of the monopole. The direct Cherenkov emission from a single muon is also shown as a comparison reference (green line).}
\label{lightyield}
\end{figure}

In this study, the KYG model is utilized for MMs cross-section with matter,
whereas the previous analyses \cite{ANTARES:2017qjw}, \cite{ANTARES:2011rzb} used the Mott model. The difference in the photon yields by a MM (the number of expected delta-rays as a function of $\beta$) derived with the two models is shown in Fig.\ref{lightyield}. The analysis is also improved by a "run-by-run" simulation strategy \cite{ANTARES:2020bhr}, that treats each data run individually, allowing for an accurate reproduction of the data taking conditions. This change is motivated by the fact that the KYG model derives from a more accurate estimation of the cross-section of MMs with matter. Since the limits set by the IceCube collaboration \cite{IceCube:2015agw} also used the KYG model, this choice allows for a more coherent comparison between the results obtained with the two experiments.

\section{Monte Carlo simulations}

For this analysis, a dedicated Monte Carlo (MC) production that includes MMs based on the KYG model for the signal has been used. The background consisting of atmospheric muons and neutrinos.
In order to take into account the variation of environmental conditions in sea water, which affects data acquisition and the optical module efficiencies, the production of simulated files containing signal and background events is performed based on run-by-run simulation. This approach addresses each run of data separately by considering its real conditions of acquisition.

The simulation of up-going MMs is carried out using 10 equally spaced velocity ranges in the interval of $\beta$ $\in$ $[0.5500, 0.9950]$.
It relies on a package which is adapted on generators used to simulate the passage of muons in the detector \cite{KUDRYAVTSEV2009339}. For each velocity range, a total of 500 MM tracks per run are generated uniformly in the surface of a cylindrical volume around the detector and according to the detector acceptance \cite{ANTARES:2020bhr}. 
The detector's response to MM signals, together with the emission, propagation and detection of direct and delta-ray Cherenkov light are then simulated, with the photon wavelength ranging between 300 and 600 nm to match the ANTARES photomultipliers sensitive range.

The background consists mainly of up-going  muons  resulting from the interaction of atmospheric neutrinos and down-going atmospheric muons  mis-reconstructed as up-going tracks.
The generator MUPAGE \cite{Bazzotti:2009zz} is used to simulate the atmospheric muons based on the parameterization of the direction and energy distributions of  under-water muons taking into account the muon  bundle  multiplicity. 


\section{Trigger and Reconstruction}
The events considered in this analysis must fulfill the conditions applied by the ANTARES triggers \cite{ANTARES:2006gto}, which are based on local coincidences defined as the occurrence of either two hits on two separate optical modules of a single storey within 20 ns, or one single hit of large amplitude (more than 3 photo-electrons). For this analysis a software trigger is defined as a combination of two local coincidences in adjacent or next-to-adjacent storeys within 100 ns or 200 ns, respectively. 

\begin{figure}[H]
\centering
\includegraphics[width=1.\linewidth]{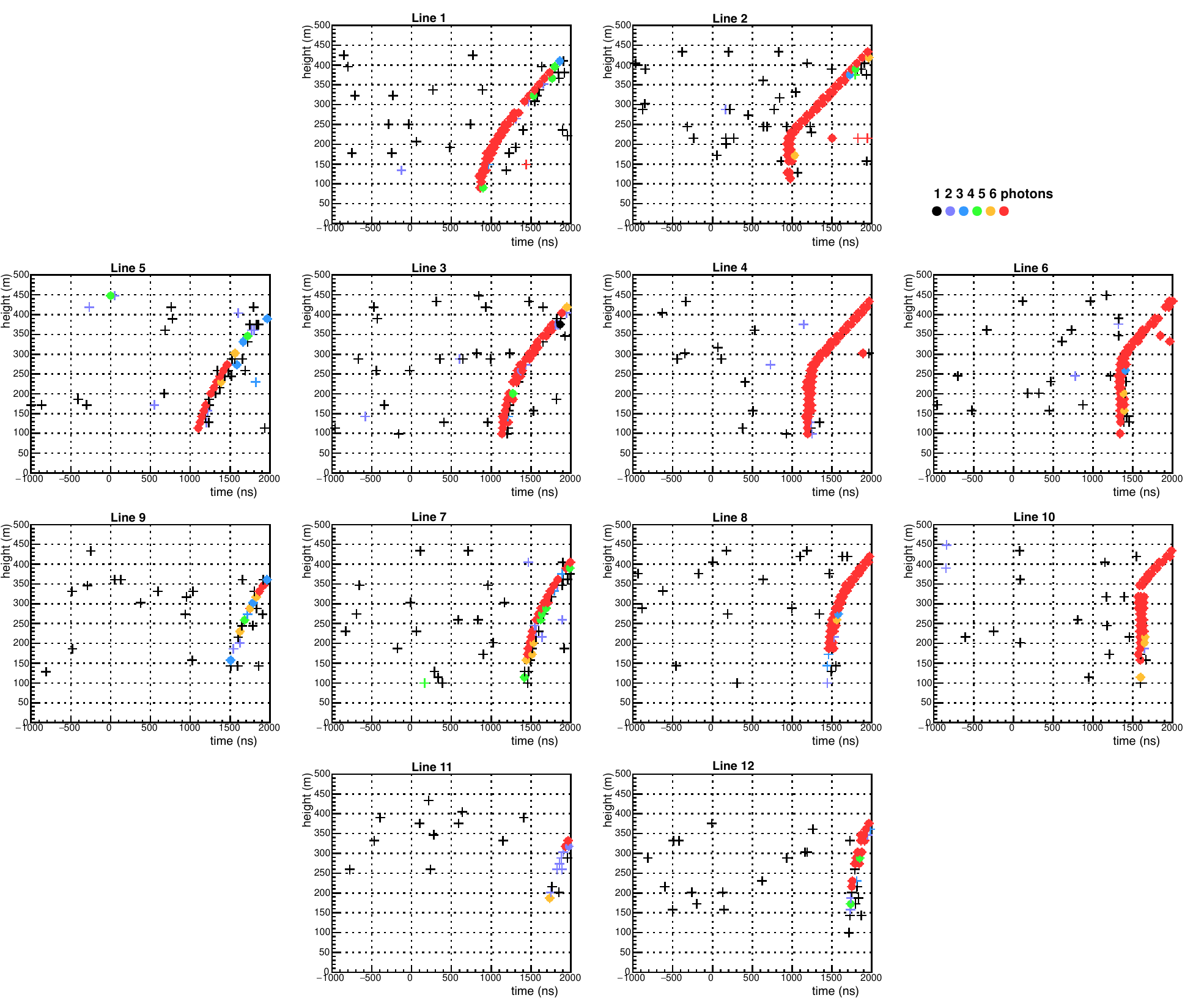}
\caption{Event display of a simulated magnetic monopole in the range $\beta\in$[0.9950, 0.9950] passing through the ANTARES telescope after traversing the Earth (up-going event). Each individual graph represents one detector line, the octagonal arrangement corresponds approximately to their layout on the sea floor and for each line the detected photons are given as function of their arrival time (x-axis) and height above the sea floor (y-axis), their amplitude is color coded as well. The MM signals (red hyperbolae) are clearly distinguishable from background photons (black crosses).}
\label{pmt}
\end{figure}

In order to reconstruct the passage of a MM in the detector, a modified version of the fast tracking algorithm \cite{ANTARES:2011vtx} is used. The algorithm searches for a straight line (i.e, a track-like event) compatible with the large amplitude hit positions and times under the Cherenkov photon emission hypothesis while allowing for a variable effective particle speed $v_{reco}$=$\beta_{reco}$ c. One example of simulated MM with $\beta$ $\in$ [0.9505, 0.9950[ is shown in Fig.\ref{pmt}. Two different approaches are followed in this study, depending on the velocity of the simulated MMs:

\begin{itemize}
\item Fast MMs are simulated with $\beta$ in the range [0.8170, 0.9950[, split into 4 equal intervals and are reconstructed with $\beta_{reco}$=1. In this velocity region, relativistic MMs are supposed to emit a significantly larger amount of Cherenkov light in the detector compared to muons, which is crucial to isolate the signal from the background. Referring to Fig.\ref{lightyield}, the light yield for a single muon is less than two orders of magnitude with respect to the one produced by a MM.  However, during the 10 years of analyzed data, events with muon bundles with multiplicity up to $\sim$100 are expected. In order to distinghish signal from background in this velocity range, the number of storeys with fired PMTs ($\mathrm{N_{sh}}$) is used in the reconstruction algorithm \cite{ANTARES:2011vtx}.
In Fig.\ref{pmt}, each point represents such a "storey hit", in which the center of the storey represents the position coordinates, the time considered is the time of the first hit and the charge is equal to the sum of the hit charges. $\mathrm{N_{sh}}$ is roughly proportional to the amount of light emitted by the particle. A large value of this quantity selects candidate MMs, which produce much more light compared to other background particles (muons and neutrinos) reaching the deep detector location.

\item Slower MMs are simulated with $\beta$ in the interval [0.5500, 0.8170[. The events are generated into 6 equally spaced intervals in this velocity range. MMs simulated with velocities within $\beta$ $\in$[0.5500, 0.8170[ are searched for with the parameter $\beta_{reco}$ used as a free parameter in the reconstruction algorithm.  The corresponding output value in the range $\beta_{reco}\in$[0.5500, 0.8170[ is used as primary cut to isolate the signal from the background, as atmospheric muons and neutrinos are mostly reconstructed as relativistic ($\beta_{reco}$ $\sim$1) particles.

\end{itemize}

\section{Event selection}
Events of interest for this study are selected by imposing several cuts to reduce the background stemming from atmospheric muons and neutrinos. The first selection parameter corresponds to the zenith angle $\theta_{reco}$ of the reconstructed track. In order to select up-going events, using the Earth as a filter, the condition $\theta_{reco} > 90^\circ$ is required. Only events reconstructed with at least 2 lines of the detector are considered to improve the quality of the event reconstruction \cite{ANTARES:2014unv}.

\begin{figure}[H]
\centering
\includegraphics[width=0.45\linewidth]{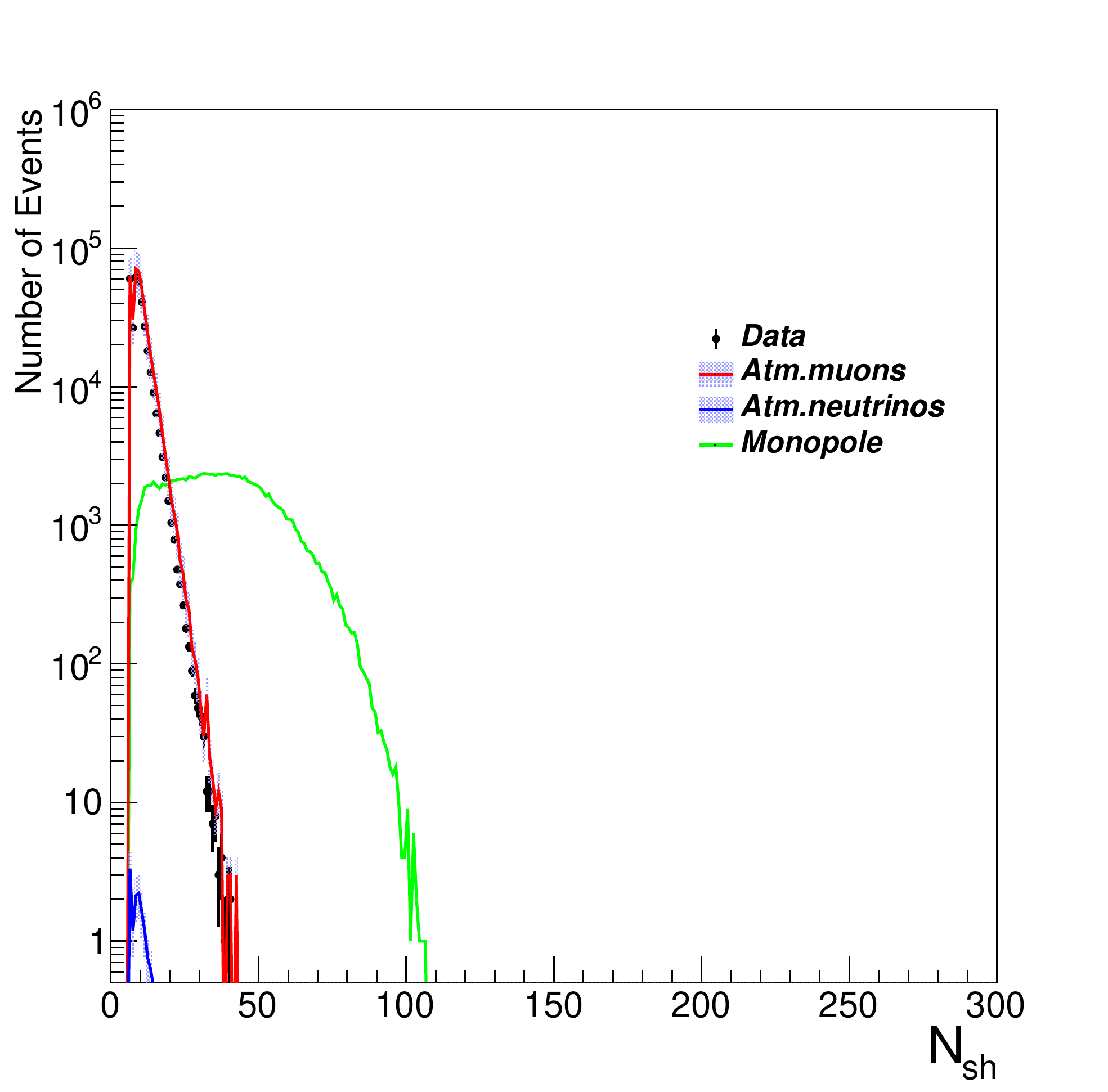}
\includegraphics[width=0.45\linewidth]{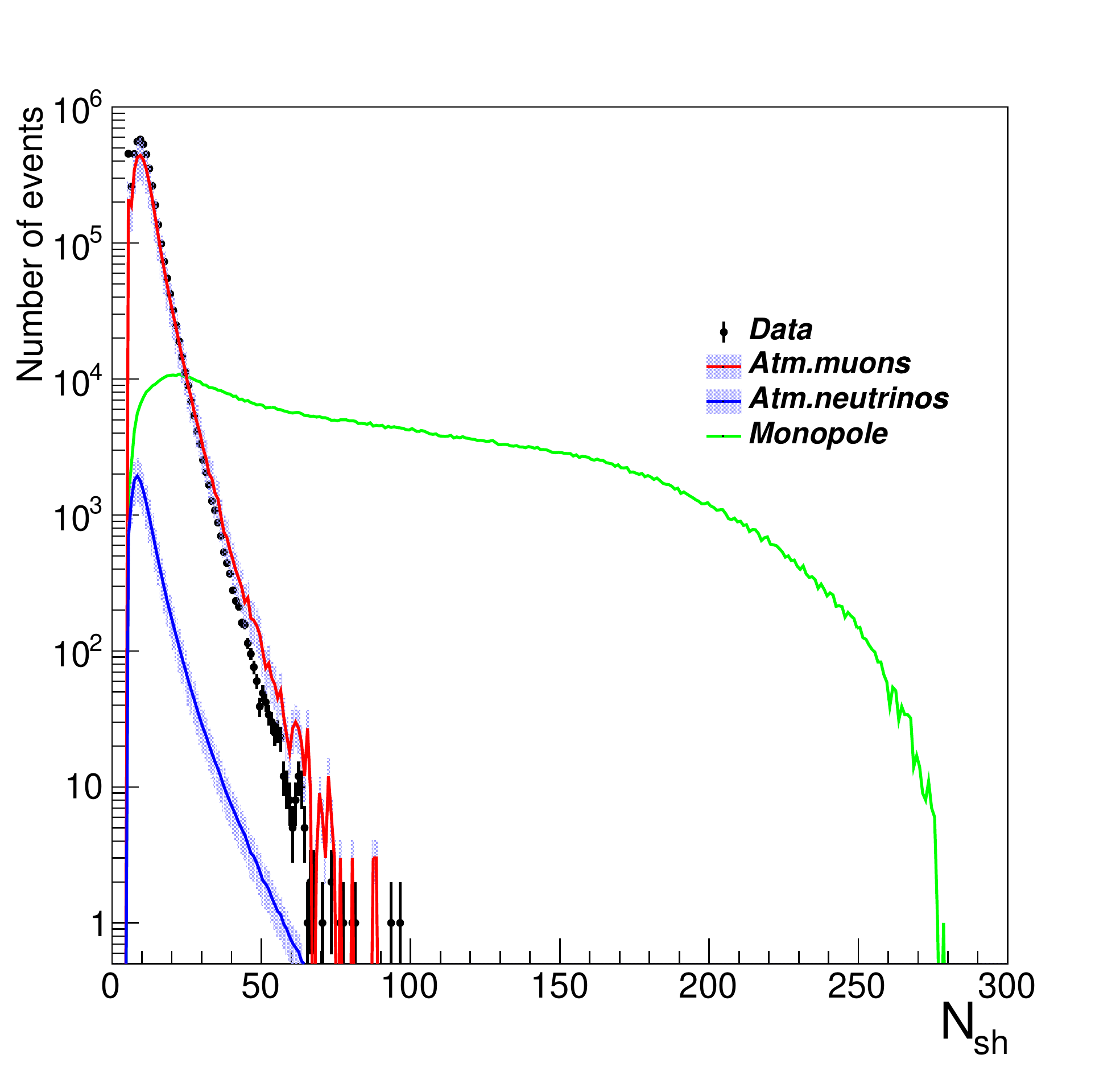}
\label{nhit}
\caption{Distribution of $\mathrm{N_{sh}}$ for atmospheric neutrinos (red histogram), atmospheric neutrinos (blue histogram) with an uncertainty band of 35\% (gray band). Data is represented in black points and MMs signal in shown in green. The plot in the left corresponds to the interval [0.5945, 0.6390[ of $\beta$ and has an additional cut (beside the initial cuts described in section 6) on $\beta_{reco}$ in [0.5945, 0.6390[, while the plot on the right corresponds equivalently to the $\beta$ range [0.8615, 0.9060[. Both plots correspond to 10 years of analyzed data.}
\end{figure}

The $\mathrm{N_{sh}} $ parameter introduced earlier is chosen as an event energy proxy in this study.
A final discriminant variable used to isolate the MM signal from the background combines the quality parameter of the track reconstruction \textit{t}$\chi{^2}$ with the brightness of the event given by $\mathrm{N_{sh}} $ reduced by the number of free parameters, $ N_{d f} $, used by the reconstruction method. This variable, denoted $\alpha$, is empirically defined as:

\begin{equation}
\alpha=\frac{t \chi^{2}}{1.3+\left(0.04 \times\left(N_{sh}-N_{d f}\right)\right)^{2}},
\end{equation}
$N_{d f}$ is equal to 6 when $\beta_{reco}$ is included in the  reconstruction, which is the case for slow MMs ($\beta_{reco}$ $\in$ [0.5500, 0.8170[), and to 5 when $\beta_{reco}$ is fixed to 1, corresponding to almost relativistic MMs.

The selection of MM events against the background is carried out under a blinded strategy to avoid any bias. The optimization of the selection parameters, $\alpha$, $\mathrm{N_{sh}} $ and $\beta_{reco}$ is made in six bins of $\beta$ in the range [0.5500, 0.8170[ and four bins in the range [0.8170, 0.9950[. In these four bins, $\beta_{reco}$ is fixed to 1.


\begin{figure}[H]
\centering
\includegraphics[width=0.48\linewidth]{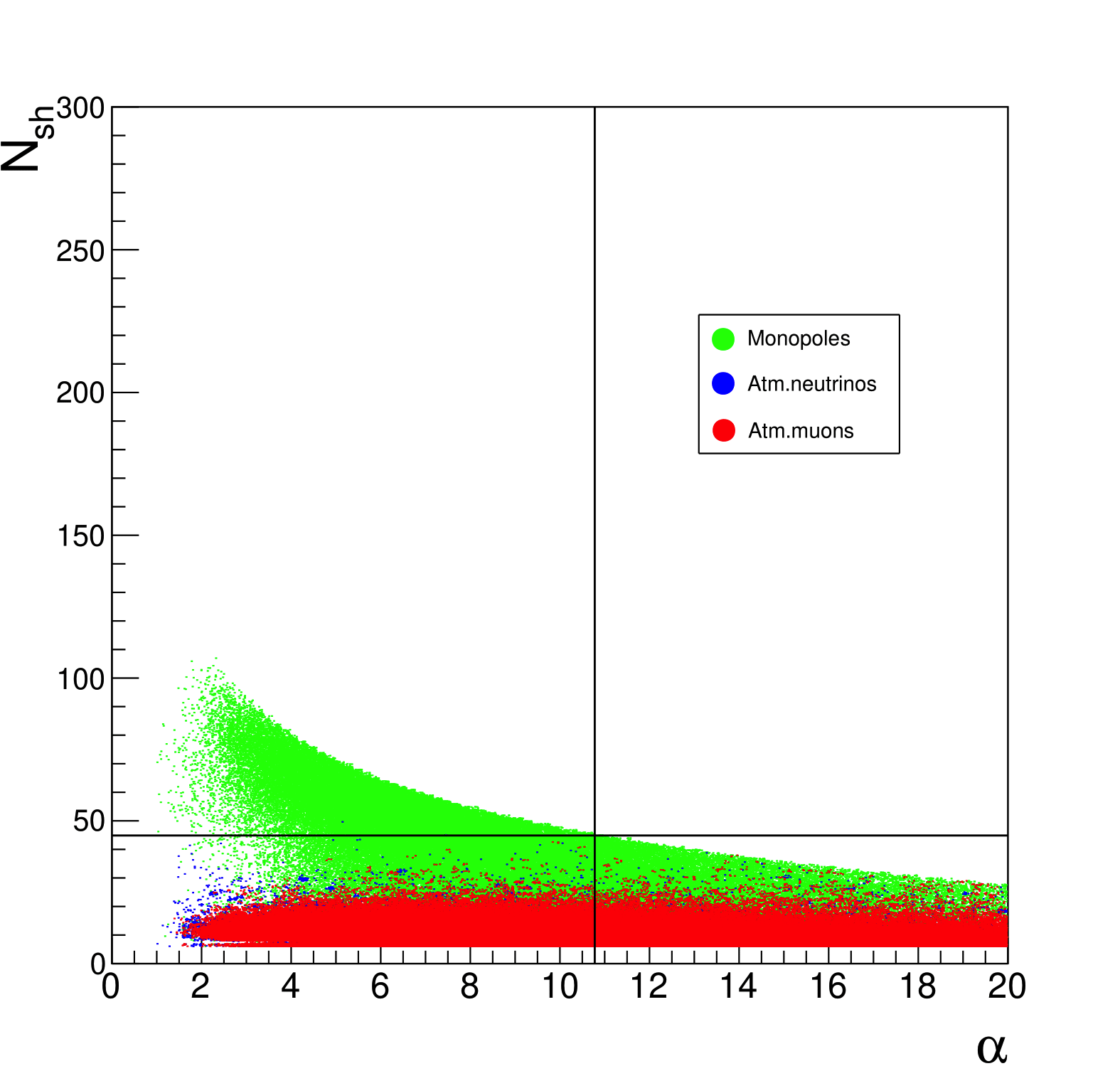}
\includegraphics[width=0.48\linewidth]{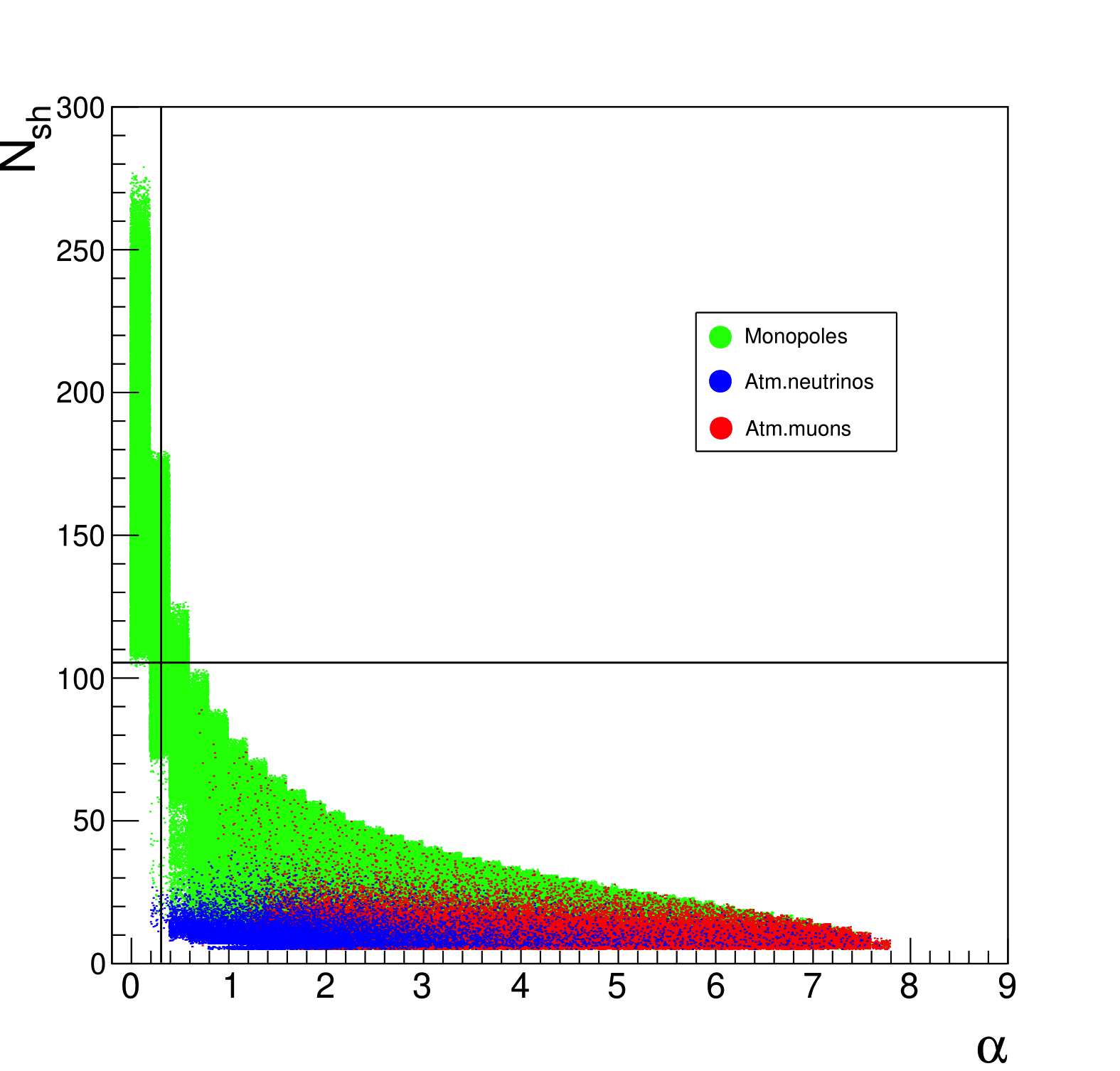}
\caption{Scatter-plot of the two variables $\alpha$ and $\mathrm{N_{sh}} $ for the MMs signal simulated with $\beta$ in the range [0.5945, 0.6390[ (left plot) with an additional cut (beside the initial cuts described in section 6) $\beta_{reco}$ $\in$ [0.5945, 0.6390[, and $\beta$ $\in$ [0.8615, 0.9060[ (right plot). The background regions consisting of atmospheric muons in red and atmospheric neutrinos in blue, are  distinguishable. The black lines indicate the optimized cuts. Both plots correspond to 10 years of analyzed data.}
\label{scatter}
\end{figure}
Fig.\ref{scatter} shows the event distribution of $\alpha$ versus $\mathrm{N_{sh}} $ for MMs simulated in the $\beta$ ranges [0.5945, 0.6390[ and [0.8615, 0.9060[. The MMs signal can be distinguished from the background (superimposed on the signal in the figure) by applying cuts on $\alpha$ and $\mathrm{N_{sh}}$.

\begin{figure}
\centering
\includegraphics[width=0.77\linewidth]{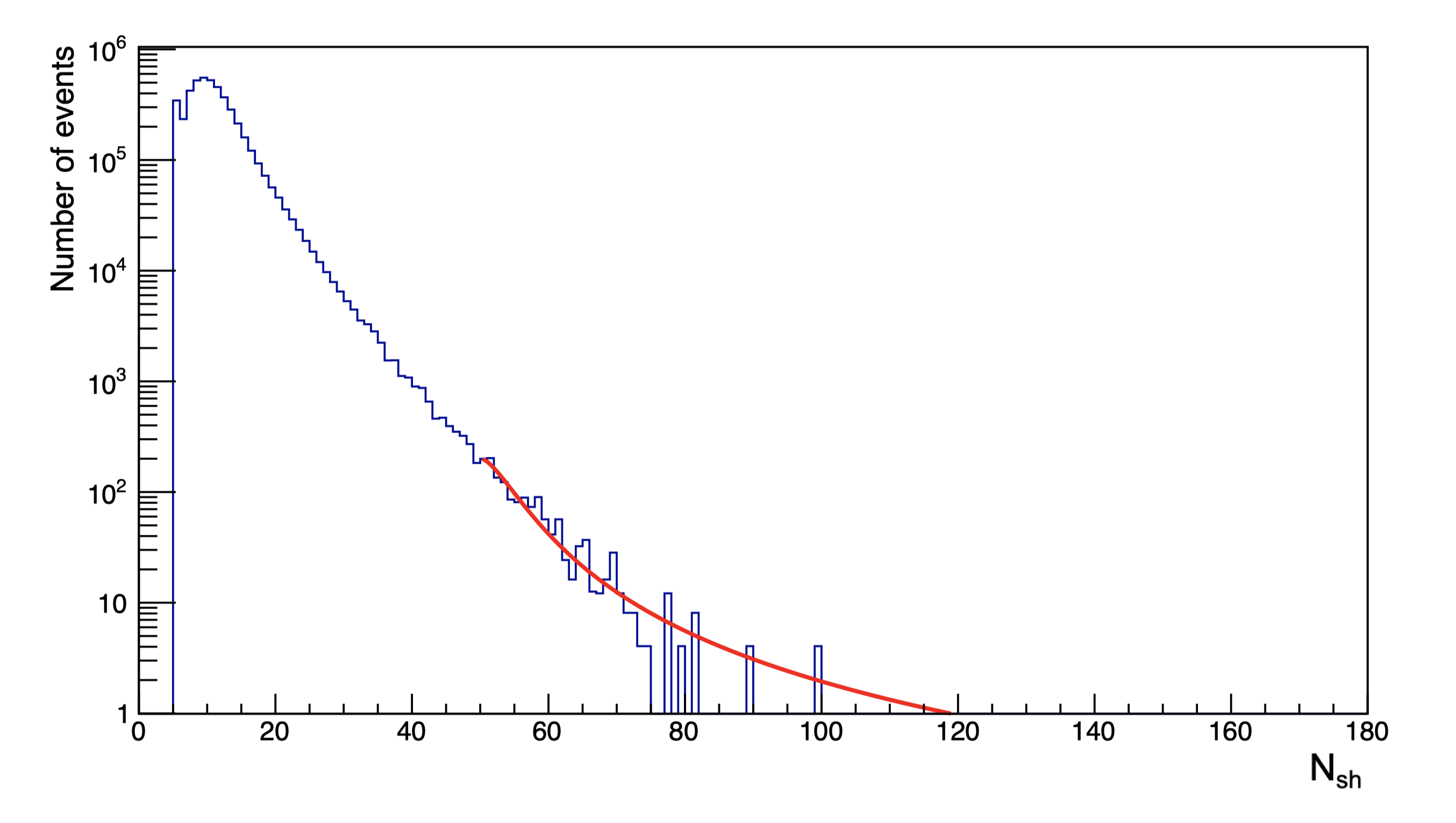}
\caption{Distribution of $\mathrm{N_{sh}} $  for simulated atmospheric muons. The red line represents the extrapolation in the signal region using a Landau function. The extrapolation is taken into account for the calculation of the the sensitivity. The plot corresponds to 10 years of analyzed data.}
\label{nhitexp}
\end{figure}

To compensate for the lack of statistics in the $\mathrm{N_{sh}} $  distribution for the simulated sample of atmospheric muons, an extrapolation is performed in the signal region, by fitting the histogram with a Landau distribution as seen in Fig.\ref{nhitexp}.
The total number of background events used to calculate the sensitivity includes the contribution of this extrapolation.

To obtain the best sensitivity, the Model Rejection Factor is optimized for each velocity interval by relying on $\alpha$ and $\mathrm{N_{sh}} $ cuts. The selection efficiency for signal events in the ten intervals of $\beta$ after applying the cuts on $\alpha$, $\mathrm{N_{sh}} $ and $\beta_{reco}$ (the latter, applied in the six bins of lower velocities) ranges from 16\% to 51\%.

\begin{figure}
\centering
\includegraphics[width=0.85\linewidth]{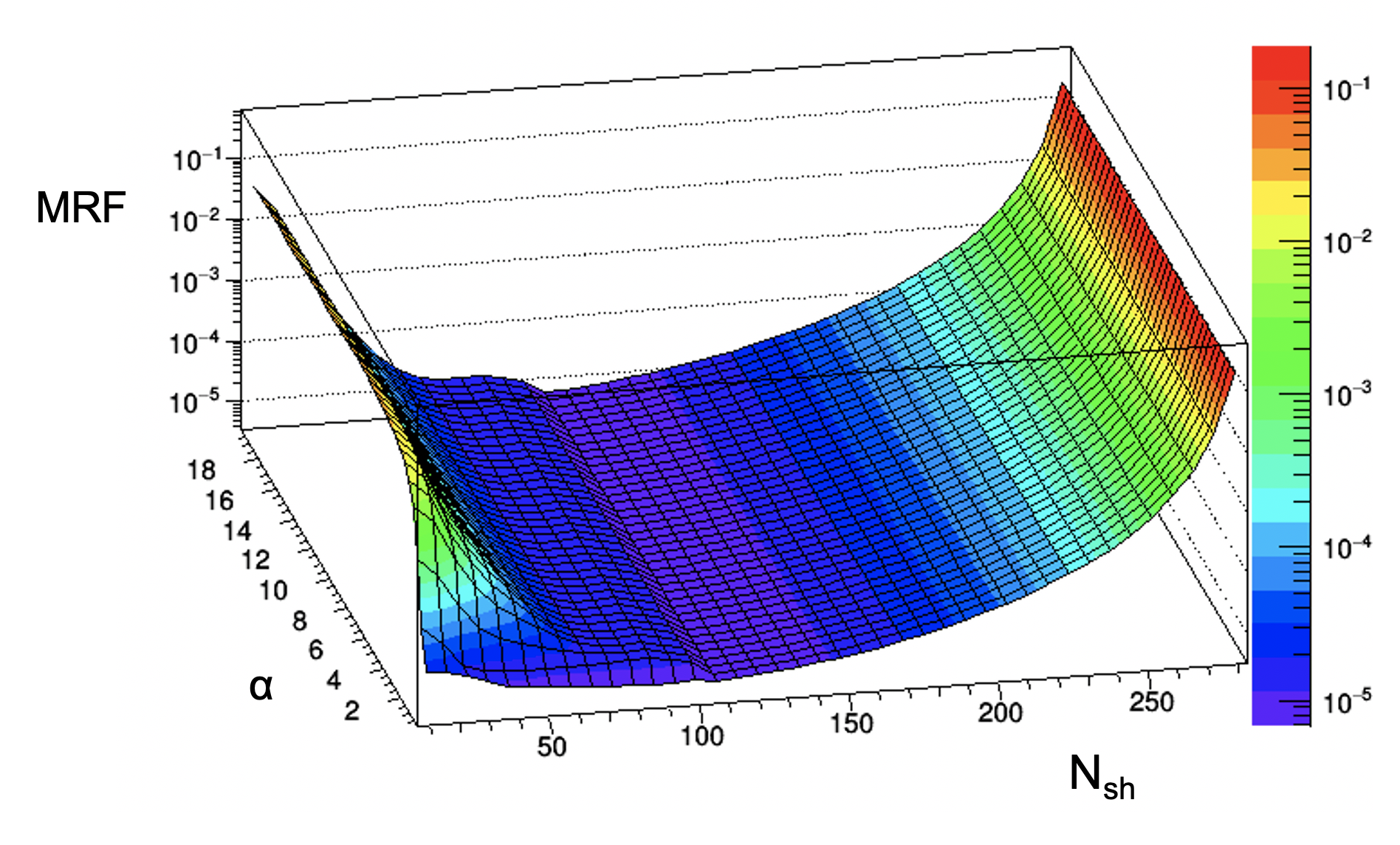}
\caption{The model rejection factor MRF as a function of $\alpha$ and $\mathrm{N_{sh}} $  cuts represented here for $\beta_{reco}$ $\in$ [0.7725, 0.8170[ as an example.}
\label{MRF}
\end{figure}

The $90 \%$ C.L. sensitivity $S_{90 \%}$ is calculated with the Feldman-Cousins \cite{Feldman:1997qc} formula, considering events which follow a Poisonnian distribution:

\begin{equation}
\centering
    S_{90 \%} [\mathrm{cm}^{-2} \:\mathrm{s}^{-1} \: \mathrm{sr}^{-1}]
    =\frac{\bar{\mu}_{90}\left(n_{b}\right)}{A_{e f f}\left[\mathrm{cm}^{2}\; \mathrm{sr}\right] \times T[\mathrm{s}]} 
    \:,
\end{equation}

\noindent where \textit{T} is the duration of data taking,  $n_{b}$ representing the number of expected background events in the 90\% C.L. interval ($\mu_{90}$), $\bar{\mu}_{90}$ and $A_{e f f}$ are defined as:
\begin{equation}
\centering
\bar{\mu}_{90}\left(n_{b}\right)=\sum_{n_{o b s}=0}^{\infty} \mu_{90} \frac{n_{b}^{n_{o b s}}}{n_{o b s} !} e^{-n_{b}} ,
\end{equation}

\begin{equation}
\centering
A_{e f f}=\frac{n_{M M}}{\Phi_{M M}} ,
\end{equation}

\noindent where $\ n_{M M}$ is the number of MMs remaining after cuts, $\Phi_{MM}[\mathrm{cm}^{-2} \, \mathrm{sr}^{-1}]$ represents the flux of generated MMs and $n_{obs}$ is the total number of observed events.
The Model Rejection Factor technique consists in varying the cuts until the minimum flux of Rejection Factor (RF) is found, which coincides with the best sensitivity (see Fig.\ref{MRF}).
After the optimization of the rejection factor RF, the sensitivity at 90\% C.L. is derived using the Feldman-Cousins method.

\section{Results}
The upper limits obtained corresponding to 10 years of data, as well as the cuts and the number of events remaining in each $\beta$ region, are summarized in Table \ref{tab:my_label}.

\begin{figure}[H]
\centering
\includegraphics[width=0.9\linewidth]{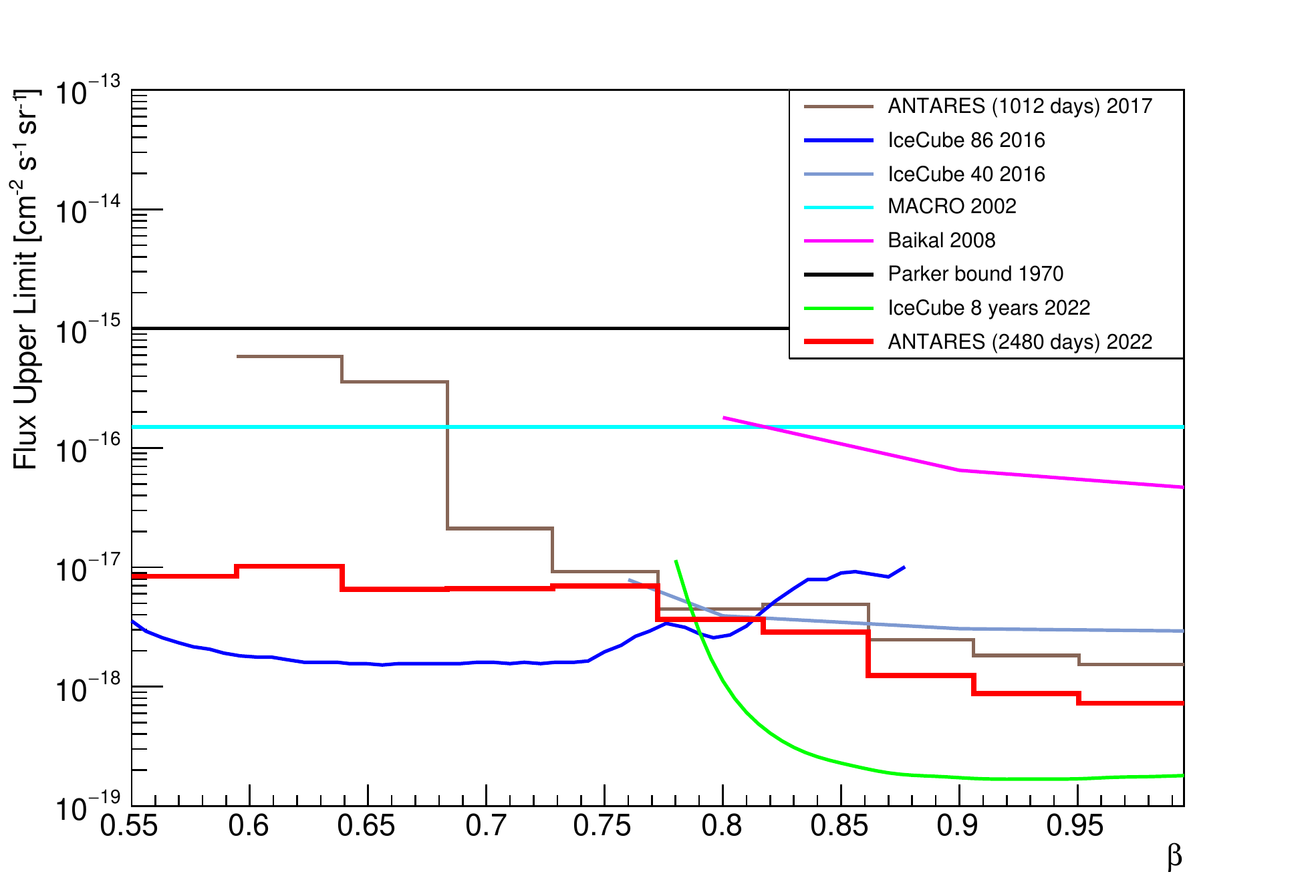}
\caption{ANTARES  90\%  C.L upper limit on the flux for MMs corresponding to 10 years of analyzed data (2480 days, red line) compared to other experiments: ANTARES previous upper limit on the flux (brown line \cite{ANTARES:2017qjw}), IceCube (blue \cite{IceCube:2015agw} and green lines \cite{IceCube:2021eye}), MACRO  (cyan line \cite{MACRO:2002jdv}) and Baikal (magenta line \cite{BAIKAL:2007kno}), as well as the theoretical Parker bound (black line \cite{Parker:1970xv}).}
\label{limit}
\end{figure}

After applying the cuts on the totality of the data taken, no event survived the selection. Fig.\ref{limit} shows the obtained ANTARES upper limit on the flux for Magnetic Monopoles, taking into account the full period of 2480 days of data taking, compared to the upper limits on the flux found by the other experiments, and including upper limits  of the previous MM analysis  (1012 days) carried out with the ANTARES telescope.

\begin{table}[H]
    \centering
\begin{center}
\scalebox{0.70}{
\begin{tabular}{|c|c|c|c|c|c|c|c|c|}
\hline
\multirow{2}{*}{$\beta$ Interval} & \multirow{2}{*}{$\beta_{reco}$} & \multirow{2}{*}{$\beta$ cut } & \multirow{2}{*}{$\alpha$ cut} & \multirow{2}{*}{$\mathrm{N_{sh}} $ cut }  &\multicolumn{1}{c|}{Expected}  & \multicolumn{1}{c|}{Observed}& \multicolumn{1}{c|}{Flux upper limit}  \\
& & &  &  & \multicolumn{1}{c|}{background} &  events  & \multicolumn{1}{c|}{$\left[\mathrm{cm}^{-2} \; \mathrm{s}^{-1}\; \mathrm{sr}^{-1}\right]$}\\\hline \hline 
[0.5500, 0.5945[  & Fitted & [0.5500, 0.5945[ &$<$ 12.4 & $\geq 41 $& 5 $\times10^{-5}$ & 0 & 8.4 $\times10^{-18}$  
\\\hline 
[0.5945, 0.6390[ & Fitted & [0.5945, 0.6390[ &$<$ 10.8 & $\geq 45 $& 2 $\times10^{-5}$  & 0  & 1.0 $\times10^{-17}$   
\\\hline 
[0.6390, 0.6835[ & Fitted & [0.6390, 0.6835[ &$<$ 8.8 & $\geq 51 $& 3 $\times10^{-4}$  & 0  & 6.5 $\times10^{-18}$   
\\\hline  
[0.6835, 0.7280[  & Fitted & [0.6835, 0.7280[ &$<$ 5.2 & $\geq 68$ & 2 $\times10^{-4}$  & 0 & 6.7 $\times10^{-18}$   
\\\hline  
[0.7280, 0.7725[ & Fitted & [0.7280, 0.7725[ &$<$ 3.6 & $\geq 85$ & 5 $\times10^{-4}$  & 0  & 7.0 $\times10^{-18}$   
 \\\hline  
[0.7725, 0.8170[ & Fitted & [0.7725, 0.8170[ &$<$ 2.6 & $\geq 86$ & 8 $\times10^{-4}$   & 0  & 3.7 $\times10^{-18}$   
 \\\hline \hline
[0.8170, 0.8615[ & 1 & - &$<$ 0.6 & $\geq 102 $ & 0.29  & 0  &   2.8 $\times10^{-18}$   
 \\\hline  
[0.8615, 0.9060[ & 1 & - &$<$ 0.3 & $\geq 105  $& 0.18 & 0  & 1.2 $\times10^{-18}$   
 \\\hline  
[0.9060, 0.9505[ & 1 & - & $<$ 0.3 & $\geq 105 $ & 0.18 & 0  &  8.8 $\times10^{-19}$   
 \\\hline 
[0.9505, 0.9950] & 1 & - & $<$ 0.3 & $\geq 105$  & 0.18  & 0  & 7.3 $\times10^{-19}$    \\\hline  
\end{tabular}
}
\end{center}

    \caption{The optimized cuts, the number of background events remaining after cuts, the number of observed events remaining after the cuts and the upper limit on the flux obtained in each $\beta$ range,  for the full analyzed data sample corresponding to 10 years live time.}
    \label{tab:my_label}
\end{table}

Above the Cherenkov threshold ($\beta$=0.76), where the emission of Cherenkov light is direct and the impact of the cross-section models can be neglected, the improvement in the upper limit on the flux for MMs compared to the results found in the previous MMs search (ANTARES limit \cite{ANTARES:2017qjw} 1012 days) is mainly due to the increase in the statistics, that is data taking time. Below the Cherenkov threshold a much substantial improvement is observed in the sensitivity, which is also a consequence of the choice of the KYG model, that can be considered more promising for the lower velocities (see Fig.\ref{lightyield}), in view of the improved description of MM cross-section and the increase in the light yield. In this region, the additional cut on $\beta_{reco}$ eliminated the majority of the background, this additional cut explains the significant drop in the number of remaining background events observed in Table \ref{tab:my_label} for $\beta<0.8170$. The background in the range of $\beta \geq 0.8170$  is dominated by atmospheric muons, while in the lower velocity range the background is dominated by atmospheric neutrinos.

\section{Conclusion}
In this work, the results of a new analysis searching for magnetic monopoles with velocities ranging in the interval of $\beta$ $\in$ [0.5500,0.9950[ crossing the neutrino telescope is presented, using an optimized simulation strategy of MMs based on the KYG model, and a larger exposure.
A Model Rejection Factor method, relying on the cuts on observable parameters, as the number of hits in the detector and the quality of the reconstructed event, has been employed to optimize the sensitivity for each of the 10 intervals of $\beta$ considered.
After the analysis of the full data sample (2480 days), no event survived the selection and upper limits on the flux are set for each of the 10 intervals.
The choice of the KYG model for the MM cross-section with matter led to the improvement in the upper limit on the flux for low velocities with respect to ANTARES previous result, as well as the extra cut that is applied in this region, which allowed a better background rejection.
The upper limit on the flux obtained in this analysis is between  $7.3\times10^{-19}$ and $1.0\times10^{-17}$ ${\rm cm^{-2} \;s^{-1} \;sr^{-1}}$, and  holds for MMs with mass $\mathrm M \gtrsim 10^{11}\; \mathrm{GeV/c^2}$, due to the requirement to cross the Earth diameter \cite{Spurio:2019oaq}, which can be considered a competitive result, in particular in view of the modest size of ANTARES with respect to IceCube.

\section*{\textbf{Acknowledgements}}
The authors acknowledge the financial support of the funding agencies:
Centre National de la Recherche Scientifique (CNRS), Commissariat \`a l'\'ener\-gie atomique et aux \'energies alternatives (CEA), Commission Europ\'eenne (FEDER fund and Marie Curie Program), Institut Universitaire de France (IUF), LabEx UnivEarthS\: (ANR-10-LABX-0023 and ANR-18-IDEX-0001),
\;R\'egion \^Ile-de-France (DIM-ACAV), R\'egion Alsace (contrat CPER), R\'egion Provence-Alpes-C\^ote d'Azur, D\'e\-par\-tement du Var and Ville de La Seyne-sur-Mer, France;
Bundesministerium f\"ur Bildung und Forschung
(BMBF), Germany; 
Istituto Nazionale di Fisica Nucleare (INFN), Italy;
Nederlandse organisatie voor Wetenschappelijk Onderzoek (NWO), the Netherlands;
Executive Unit for Financing Higher Education, Research, Development and Innovation (UEFISCDI), Romania;
Ministerio de Ciencia, Innovaci\'{o}n, Investigaci\'{o}n y
Universidades \;(MCIU): \;Programa Estatal de Generaci\'{o}n de
Conocimiento (refs. PGC2018-096663-B-C41, -A-C42, -B-C43, -B-C44) (MCIU/FEDER), Generalitat Valenciana: \;Prometeo \;(PROMETEO\;/2020/019), \;Grisol\'{i}a\; (refs. GRISOLIA\;/2018/119,
\\
/2021/192) and GenT (refs. CIDEGENT/2018/034, \;/2019/043, /2020/049, 021/023) programs, Junta de Andaluc\'{i}a (ref. A-FQM-053-UGR18), La Caixa Foundation (ref. LCF/BQ/IN17/11620019), EU: MSC program (ref. 101025085), Spain;
Ministry of Higher Education, Scientific Research and Innovation, Morocco, and the Arab Fund for Economic and Social Development, Kuwait.
We also acknowledge the technical support of Ifremer, AIM and Foselev Marine for the sea operation and the CC-IN2P3 for the computing facilities.


\bibliography{mybibfile}

\end{document}